%% file: main-arxiv.tex
\documentclass[11pt,a4paper]{article}
\usepackage[latin1]{inputenc}
\usepackage{amsmath}
\usepackage{amsfonts}
\usepackage{amssymb}
\usepackage{graphicx}
\usepackage{psfrag}
\usepackage{fullpage}
\usepackage{hyperref}
\usepackage{multicol}

\title{\textbf{Flooding Time in Opportunistic Networks under Power Law and Exponential Inter-Contact Times}\thanks{Partially supported by the Italian MIUR-PRIN'08 project COGENT (COmputational and GamE-theoretic aspects of uncoordinated NeTworks)}}
\author{
Luca Becchetti\thanks{Dipartimento di Informatica e Sistemistica, ``Sapienza'' Universit\`a di Roma, Italy, luca.becchetti@dis.uniroma1.it}
\and Andrea Clementi\thanks{Dipartimento di Matematica, Universit\`a di Roma ``Tor Vergata'', Italy, clementi@mat.uniroma2.it}
\and Francesco Pasquale\thanks{Dipartimento di Informatica, Universit\`a di Salerno, Italy, pasquale@dia.unisa.it}
\and Giovanni Resta\thanks{Istituto di Informatica e Telematica, CNR Pisa, Italy, giovanni.resta@iit.cnr.it, paolo.santi@iit.cnr.it}
\and Paolo Santi$^\P$
\and Riccardo Silvestri\thanks{Dipartimento di Informatica, ``Sapienza'' Universit\`a di Roma, Italy, silvestri@di.uniroma1.it}
}

\newtheorem{definition}{Definition}

\newtheorem{lemma}[definition]{Lemma}

\newtheorem{theorem}[definition]{Theorem}
\newtheorem{cor}[definition]{Corollary}

\newcommand{\Prob}[1]{\mathbf{P} \left( #1 \right)}
\newcommand{\Expec}[1]{\mathbf{E} \left[ #1 \right]}
\newcommand{\proof}{\noindent\textit{Proof. }}
\newcommand{\qed}{\hspace{\stretch{1}$\square$}}

\newcommand{\virgl}{\mbox{ `` }}
\newcommand{\virgr}{\mbox{ '' }}

\newcommand{\sW}{{W}}

\newcommand{\sE}{\mathcal{E}}

\newcommand{\HC}{\text{\textsc{HC}}}
\newcommand{\HD}{\text{\textsc{HD}}}
\newcommand{\NC}{\text{\textsc{NC}}}
\newcommand{\ND}{\text{\textsc{ND}}}
\newcommand{\ph}{\alpha}
\newcommand{\pl}{\gamma}

\begin{document}
\maketitle
\begin{abstract}
\input{./abstract.tex}
\end{abstract}

\section{Introduction}
\input{./trunk/intro.tex}

\section{Related works and contributions}
\input{./trunk/related.tex}

\section{The Home-MEG model: definitions and validation}\label{sec:fitting}
\input{./trunk/model-arxiv.tex}

\section{Flooding Time of Home-MEGs}\label{sec:flooding}
\input{./trunk/flooding-arxiv.tex}

\section{Conclusions}\label{sec:concl}
\input{./trunk/conclusions.tex}

\bibliographystyle{plain}
\bibliography{dynets}

\end{document}

%% file: abstract.tex
Performance bounds for opportunistic networks have been derived in a number of recent papers for several key quantities, such as the expected delivery time of a unicast message, or the flooding time (a measure of how fast information spreads). However, to the best of our knowledge, none of the existing results is derived under a mobility model which is able to reproduce the power law+exponential tail dichotomy of the pairwise node inter-contact time distribution which has been observed in    traces of several real opportunistic networks.

The contributions of this paper are two-fold: first, we present a simple pairwise contact model  -- called the Home-MEG model -- for opportunistic networks based on the observation made in previous work that pairs of nodes in the network tend to meet in very few, selected locations (home locations); this contact model is shown to be able to faithfully reproduce the power law+exponential tail dichotomy of inter-contact time. Second, we use the Home-MEG model to analyze flooding time in opportunistic networks, presenting asymptotic  bounds on flooding time that assume different initial conditions for the existence of opportunistic links. 

Finally, our bounds provide some analytical evidences that the speed of information spreading in opportunistic networks can be much faster than that predicted by simple geometric mobility models.

%% file: trunk/intro.tex
Opportunistic networks are a special class of mobile ad hoc networks where node density is so low that the network is disconnected at virtually any time. Communication is possible even in such a challenging environment by exploiting the so-called {\em store-carry-and-forward} mechanism,
according to which a packet is {\em stored} in node $A$'s buffer and {\em carried} around by the node till a communication opportunity with another node $B$ arises (whence the name of this class of networks); at this point, the packet can be {\em forwarded} from $A$ to $B$, and the process is repeated until the packet is eventually delivered to the destination. 

Opportunistic networks are receiving increasing attention in the research community, since many emerging application scenarios can be considered as instances of opportunistic networks. This is the case, for instance, of vehicular networks (at least, when traffic density is medium to low), of certain types of mobile sensor networks, and of pocket switched networks. The latter type of network is formed by powerful handheld devices -- able to establish direct wireless communication links through, e.g., a WiFi interface -- carried around by humans in their everyday life. 

A crucial aspect when attempting to analyze opportunistic network performance is mobility modeling. In fact, node mobility is a crucial communication means in opportunistic networks, and different assumptions about node mobility patterns might lead to different conclusions about network performance. Thus, credible conclusions about network performance for what concerns, e.g., flooding time, can be drawn only if the mobility model used in the analysis closely resembles mobility features observed in real-world mobility traces that are available in the literature -- see, e.g., those collected in \cite{crawdad}. In particular, the mobility feature that most profoundly impacts opportunistic network performance is the {\em inter-contact time}, defined as the time elapsing between two consecutive encounters of any two nodes $A$ and $B$ (see Section  \ref{sec:fitting}  for a formal definition).

It has recently been observed that the inter-contact time distribution in real-world mobility traces displays a dichotomy: it initially obeys a power law, but it has an exponential tail for relatively large time values \cite{Kara}. This so-called power law+exponential tail dichotomy of inter-contact time distribution is now commonly accepted in the community as a distinguishing feature of human mobility. Another interesting finding of \cite{Kara} is that the inter-contact time between a pair of nodes in the network is positively correlated with the return time to some  ``home locations". In other words, a specific pair of nodes tends to repeatedly and quite regularly meet in a few selected locations within the movement area, which explains the exponential decay in the tail of the inter-contact time distribution. 

Recently, several mobility models aimed at reproducing the power law+exponential tail dichotomy of inter-contact time have been introduced in the literature \cite{Boldrini,Hossmann,Hsu,Lee,Mei,Musolesi}, most of which are based on the notion of home location first introduced in \cite{Kara}. However, to our best knowledge, existing results on asymptotic performance bounds for opportunistic networks are based on much simpler models, which are not able to reproduce the power law+exponential tail dichotomy of inter-contact time observed in \cite{Kara} -- see next section. Thus, the problem of analyzing opportunistic network performance based on a mobility model able to reproduce this dichotomy is still open.



In this paper, for the first time, we present a {\em theoretical} analysis of opportunistic network performance based on a pair-wise contact model (which can be thought of as an abstract mobility model) able to reproduce the power law+exponential tail dichotomy of inter-contact time, thus addressing the above mentioned open problem. More specifically, we first introduce a simple, Markovian model of communication opportunities between pair of nodes, based on the observation made in \cite{Kara} that nodes tend to meet at few selected locations (homes), which explains the name of the model -- Home-Markovian Evolving Graph (Home-MEG, in short). Next, we validate the Home-MEG model against several real-world data traces, and show that it is able to closely reproduce the power law+exponential tail dichotomy of inter-contact times. Finally, we use the Home-MEG model to derive theoretical performance bounds on flooding time in opportunistic networks as the size $n$ of the network grows. As we shall see, our  bounds suggest a radically different performance than that predicted by previous  studies on flooding time based on geometric, random walk-like mobility models \cite{Clementi,Clementi1,PSSS10,PPPU11}: flooding might be very fast in opportunistic networks having  dichotomic inter-contact time.

Before ending this section, we want to comment about the importance of characterizing flooding performance in the asymptotic regime. The {\em flooding time} (or broadcast time) is defined as the time necessary for a node to deliver a packet to every other node in the network.
Accurate estimation of the flooding time can help researchers and network designers answer questions such as: ``Which is the time needed for a warning message issued by a vehicle to reach every other vehicle in the network?", or ``Which is the time needed for a virus generated by a node in a pocket switched network to propagate in the entire network?", and so on.
Our interest in asymptotic analysis stems from the fact that the number of nodes in opportunistic networks is likely to be very high: think about the number of vehicles forming a vehicular network in a large city, or the number of individuals forming a metropolitan-wide pocket switched network. \vspace{-3mm}

%% file: trunk/related.tex
Several mobility models have been recently introduced in the literature with the purpose of faithfully reproducing features observed in human mobility traces \cite{Boldrini,Hossmann,Hsu,Lee,Mei,Musolesi}. In fact, it has been observed that well-known mobility models such as random-waypoint and random walks cannot be directly used to faithfully reproduce human mobility, especially for what concerns the observed power law+exponential tail dichotomy of inter-contact time distribution \cite{Kara}. However, these models can be modified to (explicitly or implicitly) account for some form of social relationships between individuals, so that features observed in real-world contact traces emerge also in the traces generated by the mobility model at hand. Unfortunately, all the recently introduced models of human mobility share the property of introducing a strong inter-dependency between movement patterns of nodes in the network (e.g., nodes with strong social ties tend to have quite similar mobility patterns), which renders the analysis of asymptotic opportunistic network performance based on these models extremely difficult.

As a matter of fact, existing opportunistic network performance analyses are based on much simpler mobility models, if not on more abstract models of pair-wise contacts in which node trajectories in a geographical domain are not modeled, but only possible pair-wise contact opportunities. For instance, in \cite{Groene,Resta,Spyro,Zhang} the authors study unicast performance under the assumption of i.i.d., exponentially distributed inter-contact times between any pair of nodes in the network. Similarly, in \cite{Chain} the authors analyze unicast performance under the assumption of inter-contact times distributed according to a power law, while the analysis of \cite{Hong} is based on the assumption of a truncated power law to model inter-contact times. Simple geometric, random-walk  mobility models have been instead used to analyze flooding  performance in opportunistic networks -- see \cite{Clementi,Clementi1,Jacquet,PSSS10,PPPU11}. 

Summarizing, we currently are in a situation in which there is a quite deep understanding of the main features of human mobility, and a good number of mobility models aimed at reproducing these features have been designed. On the other hand, existing analytical  performance bounds for opportunistic networks have been derived based on much simpler mobility models, which are known not to be able to reproduce typical human mobility patterns. Thus, currently there is a gap between the state of knowledge in human mobility modeling, and what is known about opportunistic network performance bounds.

In this paper, we make a substantial step forward towards closing this gap by presenting, for the first time, asymptotic performance bounds for opportunistic networks based on a contact model which we show to be able to reproduce a salient feature of human mobility, namely, that pair-wise inter-contact times display a power law+exponential tail dichotomy. More specifically, we present bounds on flooding time, which were previously derived only using simple geometric, random-walk mobility models which are known not to be able to reproduce the above described dichotomy\footnote{Although the authors of \cite{Kara} introduce versions of random waypoint and random walk mobility able to reproduce the power law+exponential tail dichotomy, the mobility models used in the analyses of \cite{Clementi,Clementi1,Jacquet,PSSS10,PPPU11} are based on standard random waypoint/random walk models, which do not reproduce the inter-contact time dichotomy.}.

Before ending this section, we want to outline that different mobility assumptions have a profound impact on the derived asymptotic performance bounds. For instance, considering unicast performance, denoting with $E[T_n]$ the expected packet delivery time in a network with $n$ nodes, we have that $\lim_{n\to\infty}E[T_n]=c$, for some constant $c>0$, in case of exponentially distributed inter-contact times \cite{Groene,Spyro,Zhang}, while $\lim_{n\to\infty}E[T_n]=+\infty$ if inter-contact times obey a power law with sufficiently low exponent \cite{Chain}.

The bounds derived in this paper seem to suggest a similar impact of mobility assumptions on flooding performance. In fact, if we consider $n$ nodes moving with random walk-like  mobility over  a square area of side $L=\Theta(\sqrt{n})$ with  transmission range and node speed set to some positive constant, then      the time needed to complete flooding is $\Omega(\sqrt{n})$, a lower bound which has been recently shown to be almost tight \cite{PSSS10,PPPU11}. Thus, existing results suggest that flooding time grows unboundedly with $n$, at rate which is {\em polynomial} in $n$. On the contrary, our results suggest that, in a network with similar density as that considered in \cite{PSSS10,PPPU11}, flooding time grows unboundedly with $n$, but with a rate which is {\em logarithmic} in $n$ -- for a more detailed discussion on this important issue, see Section \ref{sec:flooding}. \vspace{-3mm}

%% file: trunk/model-arxiv.tex
\subsection{The Home-MEG model for a single link}
When modeling asymptotic opportunistic network performance, at least the following two options can be undertaken to model inter-contact times: $i)$ using a geometric mobility model defined in a specific geographic domain (e.g., random-walk in a unit square), and deducing pair-wise contact patterns based on node trajectories, coupled with a notion of transmission range -- see, e.g., \cite{Clementi,Clementi1,Jacquet,PSSS10,PPPU11}; or $ii)$ abstracting details of the underlying geographic domain and modelling directly inter-contact time between nodes -- see, e.g., \cite{Chain,Groene,Hong,Resta,Spyro,Zhang}. Both approaches have pros and cons. In the case of $i)$, pair-wise contact traces are by construction consistent with a reference geometry and, under that respect, can be considered realistic. On the other hand, the analysis of asymptotic performance bounds under approach $i)$ is quite complex, due to the need of deriving contact traces from an underlying mobility model. As a matter of fact, this intrinsic complexity is the reason why existing analyses based on approach $i)$ are restricted to very simple underlying mobility models such as random waypoint and random walk, which are known not to be able to reproduce the power law + exponential tail dichotomy of inter-contact time distribution typical of human mobility. Conversely, approach $ii)$ is apt for asymptotic performance analysis, since pair-wise contact patterns are modeled directly instead of being derived from an underlying mobility model. On the other hand, contact patterns under approach $ii)$ lack to refer to an underlying reference geometry, hence some aspects related to human mobility (e.g., the fact that when a node is in a certain region $R$ of the movement domain, it can come into contact only with nodes that are located in $R$ or in nearby locations) are not captured by contact models.

In this section, we introduce, to the best of our knowledge, the first contact model which: $1)$ is able to faithfully reproduce the power law+exponential tail dichotomy of inter-contact time typical of human mobility, and $2)$ can be used to derive asymptotic performance bounds for opportunistic networks. Although we admit that, as described above, contact models do not capture all aspects of human mobility, based on $1)$ we claim that the model defined herein is realistic in the sense that, differently from the models used to derive existing asymptotic performance bounds \cite{Chain,Clementi,Clementi1,Groene,Hong,Jacquet,PSSS10,PPPU11,Resta,Spyro,Zhang}, it is able to reproduce the power law+exponential tail dichotomy of inter-contact time observed in human mobility traces.

\begin{figure}[h]
\centering
\centerline{\includegraphics[width=6cm]{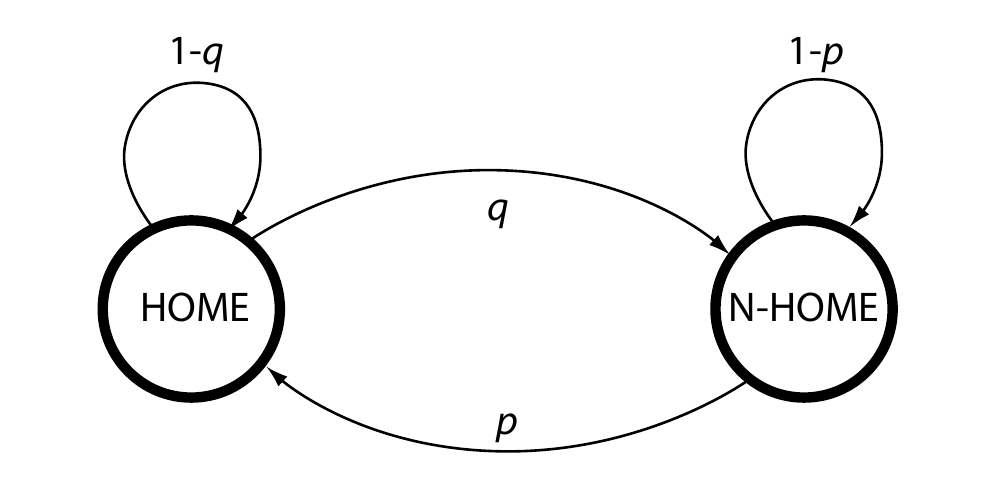}\vspace{-3mm}}
\caption{The Home-MEG model.\vspace{-12mm}}
\label{HomeMEG}
\end{figure}

As the name suggests, to increase accuracy of the generated contact traces the Home-MEG model is based on the notion of home location introduced in \cite{Kara}. 
A pictorial representation of the Home-MEG model is reported in Figure \ref{HomeMEG}. The state transition diagram reported in the figure models probability of occurrence of a wireless link between a specific pair of nodes $u,v$ in the network at discrete time $t$. We can interpret the state transition diagram reported in Figure \ref{HomeMEG} as follows. The pair of nodes $u,v$ can be in one of two states: {\em Home} state, corresponding to the situation in which both nodes $u$ and $v$ are at one of their home locations; and {\em Non-Home} state, corresponding to the complementary situation in which one of the nodes (or both) are not in a home location. We recall that, according to \cite{Kara}, any pair of nodes in an opportunistic network tend to repeatedly meet in a few locations (home locations), while meeting opportunities outside the home sites are occasional. To account for this, in the Home-MEG model the probability of establishing an {\em instantaneous} communication opportunity (i.e., a contact between $u$ and $v$) at time $t$ depends on the state of the pair: it is $\alpha$ with $0<\alpha<1$ when the pair is in state $H$, and it is $\gamma$ with $0<\gamma\le \alpha$ when the pair is in state $NH$. Finally, two further parameters of the model are the probability $q$ of making a transition from {\em Home} to {\em Non-Home} state, and the probability $p$ of making a transition from {\em Non-Home} to {\em Home} state. Summarizing, the model is fully characterized by four parameters: the state transition probabilities $p$ and $q$, and the contact opportunity probabilities $\alpha$ and $\gamma$.

The goal of the Home-MEG model is modeling {\em communication opportunities} or {\em contacts} between $u$ and $v$, where a communication opportunity is intended here as an {\em instantaneous event} during which an arbitrarily large number of messages can be exchanged. This notion of communication opportunity is consistent with a scenario in which the time step considered for analyzing information propagation within the network is relatively large, and the amount of data to be transferred between nodes is small. For instance, we can think of time steps elapsing a few minutes -- this is consistent with all real-world data collected so far \cite{Kara}; if a communication opportunity occurs between $u$ and $v$ at time $t_0$, it is very reasonable then to assume that all messages in the respective buffers can be exchanged well before the next time step $t_1$. 

Finally, notice that the Home-MEG model as defined here is similar to the Gilbert-Elliot model \cite{Elliot,Gilbert} of communication channel, which has been extensively used in the communication engineering community to estimate probability of bit error bursts on a channel. 

\subsection{Inter-contact time in the Home-MEG model}

The inter-contact time distribution in the Home-MEG model can be derived recursively as follows -- see \cite{Martelli}, where the Gilbert-Elliot model is used. In the following, $IC$ is the random variable corresponding to the number of time steps elapsing between two consecutive contact opportunities.

Let $p_H$ and $p_{NH}$ represent the stationary probabilities of finding the pair of nodes in state H and NH, respectively.
It is easy to prove that 
\[ p_H=\frac{p}{p+q} \ \mbox{ and }   \ p_{NH} =1-p_H \]
The probability $P(H|Contact)$ (respectively, $P(NH|$ $Contact)$) of finding the pair in state $H$ (respectively, $NH$), conditioned on the event that a contact occurs can be computed applying Bayes' theorem:

\begin{eqnarray}
P(H|Contact)&=&\frac{P(Contact|H)\cdot P(H)}{P(Contact)}=\nonumber\\
&=&\frac{\alpha\cdot p}{p\cdot \alpha+q\cdot \gamma}\nonumber
\end{eqnarray}
and
\[
P(NH|Contact)=\frac{\gamma\cdot q}{p \cdot \alpha + q\cdot \gamma}\nonumber
\]

The value of $P(IC=k)$, for any $k\ge 1$, is recursively defined as follows:

\[
P(IC=k)= P(H|Contact) P_{kH}+P(NH|Contact) P_{kN}
\]
where
\[
P_{1H}=(1-q)\alpha +q\gamma~, \ \
P_{1N}=(1-p)\gamma +p\alpha
\]
\[
P_{iH}=q(1-\gamma) P_{(i-1)N}+(1-q)(1-\alpha)P_{(i-1)H}
\]
\[
P_{iN}=(1-p)(1-\gamma) P_{(i-1)N}+p(1-\alpha)P_{(i-1)H}~,  \ 
  i=2,..,k \]   

\subsection{Validation}\label{subse:val}
\begin{figure*}[th]
\centering
\centerline{\includegraphics[width=5.8cm]{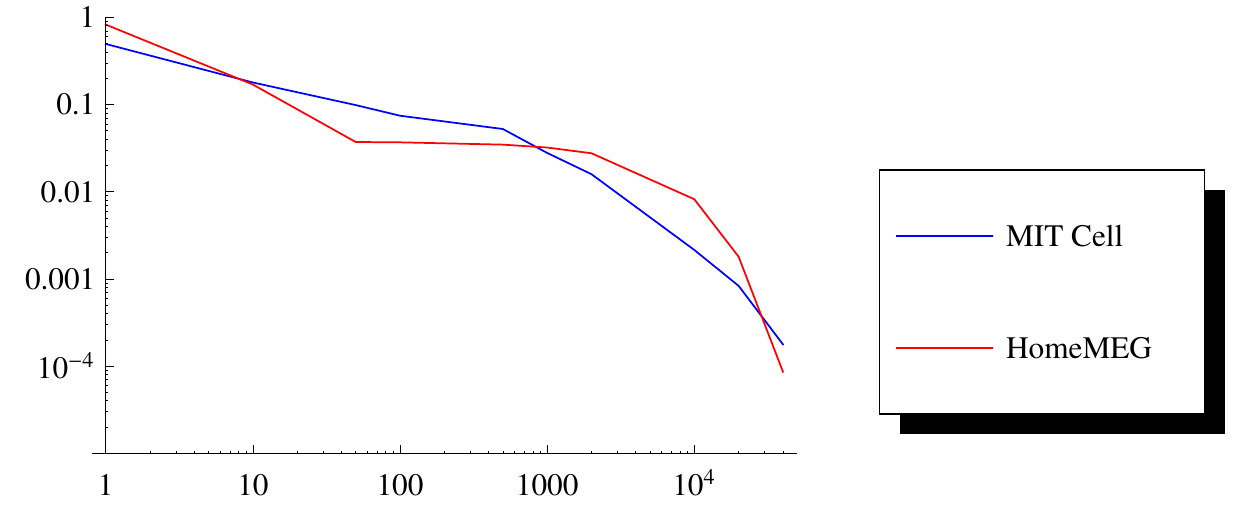}~~~~~\includegraphics[width=5.8cm]{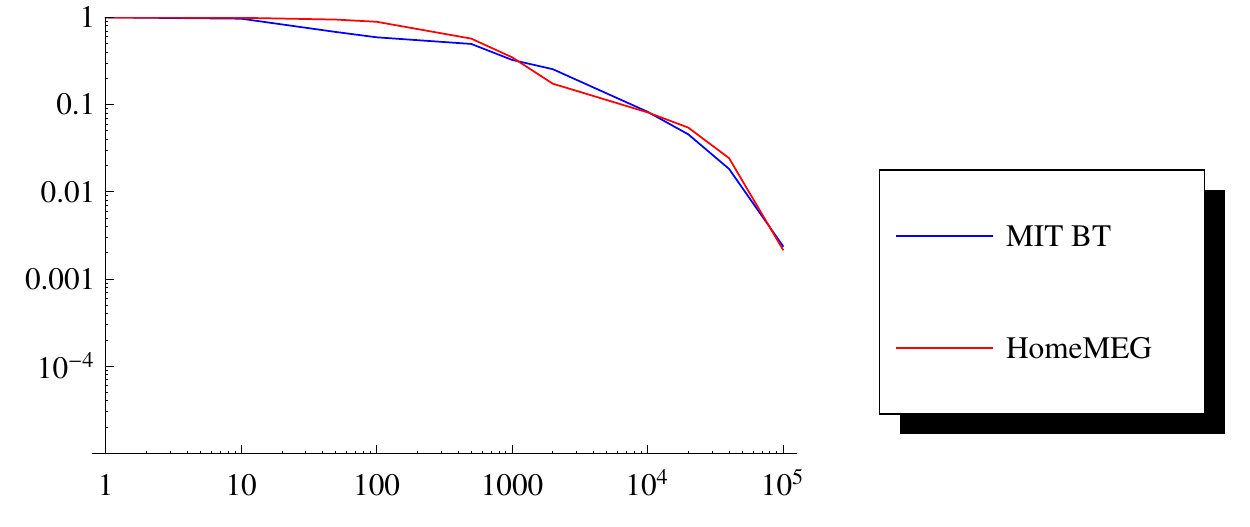}~~~~~\includegraphics[width=5.8cm]{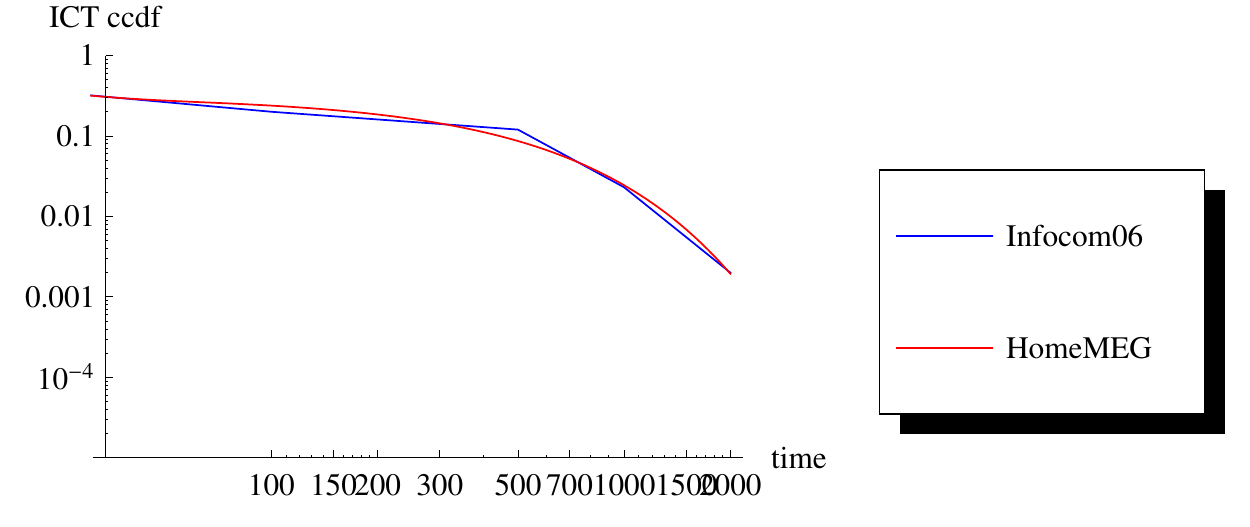}}
\centerline{\includegraphics[width=5.8cm]{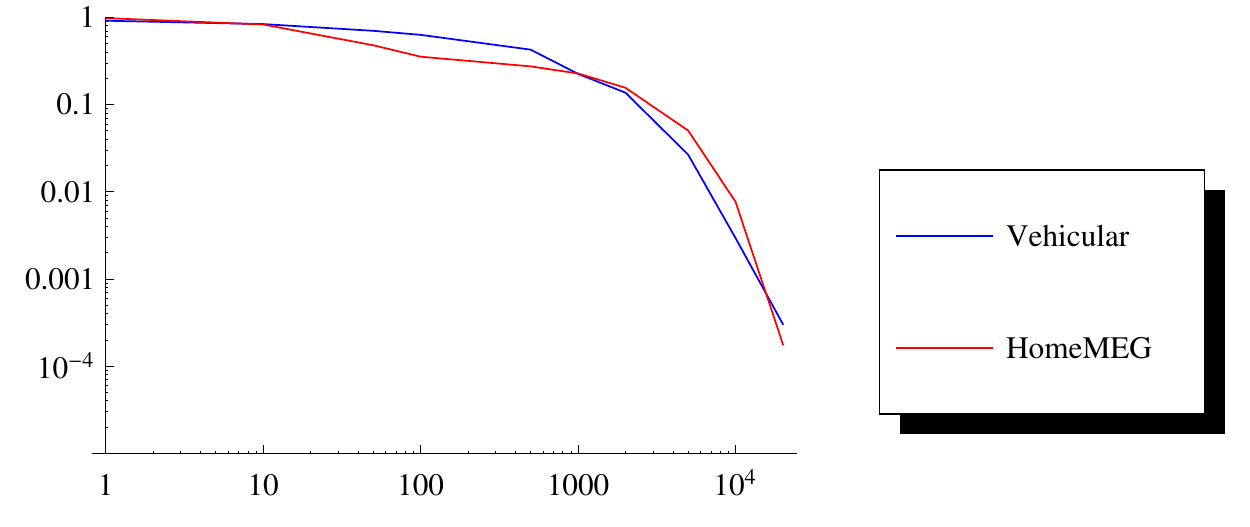}~~~~~\includegraphics[width=5.8cm]{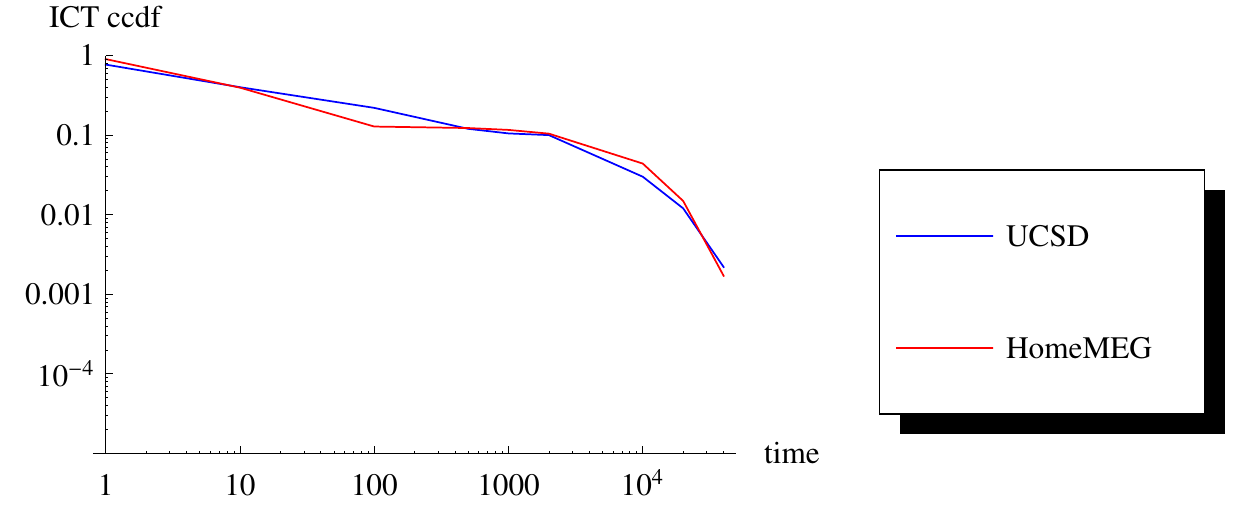}~~~~~\includegraphics[width=5.8cm]{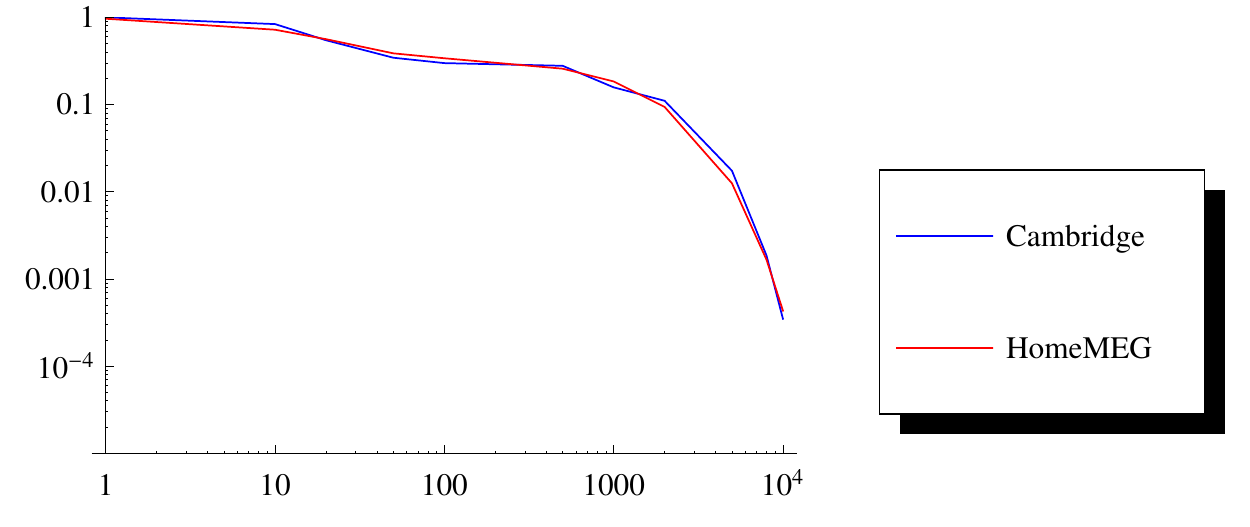}}
\caption{Best fit of the Home-MEG model (red curve) vs. six real-world data traces (blue curve).}
\label{fit!}
\end{figure*}

{\em Can we set the parameters of the Home-MEG model so that it fits real-world data traces?} To answer this question, we have considered the six data traces used in \cite{Kara}, selecting a few representative points from the corresponding complementary cumulative density function (ccdf) of the {\em aggregate} inter-contact time distribution. The main features of the considered traces are reported in Table \ref{dataTraces} (see \cite{Kara} for more details on the data traces). We have then performed an iterative search on the four-dimensional parameter space of the Home-MEG model, aiming at minimizing the mean square error (in log scale) between the data trace of reference and the ccdf produced by the Home-MEG model. The resulting best fittings for the six traces are reported in Figure \ref{fit!}. As seen from the figure, the fitting is very good, and in particular the well-known power-law + exponential tail behavior of the aggregate inter-contact time ccdf is well reproduced by the Home-MEG model. 

\begin{table}
\begin{center}
\small
\begin{tabular}{|c|c|c|c|c|}
\hline
Name & Technology & Duration & $\sharp$ devices & $\sharp$ contacts\\ \hline
MIT Cell & GSM & 16 months & 89 & 1,891,024\\
MIT BT & Bluetooth & 16 months & 89 & 114,046\\
Infocom06 & Bluetooth & 3 days & 41 & 28,216\\
Vehicular & GPS & 6 months & 196 & 9,588\\
UCSD & WiFi & 77 days & 275 & 116,383\\
Cambridge & Bluetooth & 11.5 days & 36 & 21,203\\
\hline
\end{tabular}
\caption{Main features of the considered real-world data traces.}
\label{dataTraces}
\end{center}
\end{table}

The values of the four model parameters corresponding to the best fit for the six traces are reported in Table \ref{fitValues}. In all cases, the time step in the Home-MEG model was set to 86.4 seconds. The table reports also the stationary probability $p_{H}$ of finding the system in HOME state (both nodes in one of their home locations), the ratio $\alpha/\gamma$ of link occurrence probabilities in the HOME and Non-HOME states, and the sum $p+q$ of the probabilities of making a state transition. As seen from the table, despite the different values of the four parameters for the different data traces, some general conclusions about these values can be drawn, namely: $i)$ users spend most of the time in Non-HOME state -- $p_{H}$ is consistently much smaller than $p_{NH}=1-p_{H}$; $ii)$  $\alpha/\gamma\gg 1$, i.e.,  $\alpha\gg \gamma$, which is in accordance with the intuition upon which the model has been defined; $iii)$ $p+q\ll 1$, i.e., states are persistent (otherwise stated, transition events occur seldom).

Notice that property $i)$ above is apparently in contradiction with recent characterizations of human mobility, according to which individuals tend to spend most of their time in a few locations \cite{Song}. However, we reiterate that, differently from most literature on human mobility models, the notion of home in the Home-MEG model refers to a {\em pair} of nodes, and is aimed at modeling the fact that the two nodes are {\em simultaneously} present in one of their meeting sites (home locations). Thus, the fact that a single node $u$ spends most of the time in a few locations {\em is not} in contradiction with the fact that node $u$ and, say, node $v$ are only seldom co-located in one of their meeting sites.

\begin{table*}[t]
\begin{center}
\small
\begin{tabular}{|c|c|c|c|c|c|c|c|}
\hline
Trace & $p$ & $q$ & $\alpha$ & $\gamma$ & $p_{H}$ & $\alpha/\gamma$ &  $p+q$ \\ \hline
MIT Cell & $7.5\cdot 10^{-5}$ & $3.3\cdot 10^{-3}$ & $1.8\cdot 10^{-1}$ & $7.8\cdot 10^{-3}$ &
0.022 & 2,271 & $3.4\cdot 10^{-3}$\\
MIT BT & $4.5\cdot 10^{-5}$ & $1.5\cdot 10^{-4}$ & $1.2\cdot 10^{-3}$ & $8.6\cdot 10^{-7}$ &
0.228 & 1,368 & $2.0\cdot 10^{-4}$\\
Infocom06 & $3\cdot 10^{-3}$ & $2.5\cdot 10^{-2}$ & $7\cdot 10^{-2}$ & $3\cdot 10^{-4}$ &
0.107 & 233 & $2.8\cdot 10^{-2}$\\
Vehicular & $4.1\cdot 10^{-4}$ & $7.9\cdot 10^{-3}$ & $2.1\cdot 10^{-2}$ & $7.7\cdot 10^{-5}$ &
0.049 & 275 & $8.3\cdot 10^{-3}$\\
UCSD & $1.1\cdot 10^{-4}$ & $1.3\cdot 10^{-2}$ & $10\cdot 10^{-2}$ & $1\cdot 10^{-5}$ &
0.008 & 10,000 & $1.4\cdot 10^{-2}$\\
Cambridge & $2.5\cdot 10^{-4}$ & $8.3\cdot 10^{-3}$ & $4.7\cdot 10^{-2}$ & $4.6\cdot 10^{-4}$ &
0.029 & 102 & $8.6\cdot 10^{-3}$\\
\hline
\end{tabular}
\caption{Best fit parameter values for the six data traces.}
\label{fitValues}
\end{center}\vspace{-10mm}
\end{table*}
\vspace{-4mm}
\subsection{Evolving networks and Home-MEG Model}\label{subsec:HEMEG}
The Home-MEG model can be used to describe the evolution of a network of mobile nodes that can have pairwise interactions over time.
For the purpose of analysis, we now formally re-state the Home-MEG model as a collection of independent Markov chains. Consider a time evolving graph $\mathcal{G}=([n],E_t)$, where $[n]$ is the (fixed) node set and $E_t$ is the time evolving edge set. Suppose every edge of the graph can be in four states $\{$$HC$, $HD$, $NC$, $ND$$\}$. In states $HC$ and $NC$ we say the edge \emph{exists} (and the nodes are in Home and Non-Home states, respectively), while in states $HD$ and $ND$ we say the edge \emph{does not exist}. The Markov chain over such a state space is given by the following transition matrix:

\begin{footnotesize}
\begin{equation}\label{eq:EdgeHomeMEGtm}
\left(
\begin{array}{c|c|c|c|c}
& \HC & \HD & \NC & \ND \\[1mm]
\hline
\HC & (1-q)\ph & (1-q)(1-\ph) & q \pl & q (1-\pl) \\[2mm]
\HD & (1-q) \ph & (1-q)(1-\ph) & q \pl & q (1-\pl) \\[2mm]
\NC & p \ph & p (1-\ph) & (1-p) \pl & (1-p) (1-\pl) \\[2mm]
\ND & p \ph& p (1-\ph) & (1-p)\pl  & (1-p)(1-\pl) \\
\end{array}
\right)
\end{equation}
\end{footnotesize}

\noindent Such a chain is clearly ergodic (irreducible and aperiodic) and it is easy to verify that the following state distribution is stationary
\begin{equation}\label{eq:lbprob}
\pi =
\frac{1}{p+q}\left(
\begin{array}{cccc}
\HC & \HD & \NC & \ND \\[1mm]
\hline
p \ph & p (1-\ph) & q \pl & q(1-\pl)
\end{array}
\right)
\end{equation}

\smallskip\noindent
The Home-MEG model for a network of $n$ nodes can now be formally defined as follows: 

\begin{definition}[Home-MEG]\label{HomeMEGDef}
A \emph{Home-MEG} model $\mathcal{H} = \mathcal{H} (n,p,q,\alpha,\gamma)$ on $n$ nodes, where $n \in \mathbb{N}^+$ and  $0 \leqslant  p,q,\alpha,\gamma  \leqslant 1$, is a sequence of random graphs $\mathcal{H} = \{ G_t = ([n],E_t) \;:\; t \in \mathbb{N}\}$ such that the set of edges at time $t$ is
$$
E_t =  \left\{ e \in \binom{[n]}{2} \;:\; X_t(e) \in \{HC,NC\} \right\},
$$
where $\left \{ \{X_t(e) \}_{t \in \mathbb{N}} \;:\; e \in \binom{[n]}{2} \right\}$ is a family of independent Markov chains with transition matrix as in (\ref{eq:EdgeHomeMEGtm}).
\end{definition}

Depending on the specific assumptions, the initial set of edges $E_0$, which is determined by the initial state of the Markov chains, can be arbitrary, or determined by the stationary distribution of the Markov chains (see below).

We stress again that the notion of {\em Home} state in the Home-MEG model is referred to a {\em specific} node pair, and that the $n(n-1)/2$ identical Markov chains used to model all possible links in the network are {\em independent}. The above implies that, for instance, the transitive property does not hold. In other words, the fact that the pairs of nodes $\{u,v\}$ and $\{u,w\}$ are in {\em Home} state does not imply that also pair $\{v,w\}$ is in {\em Home} state. Thus, {\em Home} state in the Home-MEG model must {\em not} be intended as a single popular location within the network where {\em all} the nodes regularly meet, but rather as an abstract location corresponding to the (few) physical locations where a specific pair of nodes regularly meet. Notice that the independence assumption in Definition \ref{HomeMEGDef} implies that meeting locations are assumed to be {\em different} for each pair of nodes. Admittedly, this is an approximation of reality, where meeting locations are likely to be shared by several pairs of nodes. However, in the previous section we have shown that, despite this approximation, the Home-MEG model is able to accurately reproduce the {\em aggregate} inter-contact time distribution observed in several real-world traces. \vspace{-2mm}

%% file: trunk/flooding-arxiv.tex
Let $\mathcal{G}$ be an evolving graph, the \emph{flooding process} in $\mathcal{G}$ proceeds as follows: Start with one single \emph{informed} node and, during the \emph{evolution} of the graph, when a non-informed node \emph{gets in touch} with an informed one, it \emph{collects} the information.  

Let $\mathcal{G} =  \{ G_t = ([n], E_t) \, : \, t \in \mathbb{N} \}$ be an evolving graph and let $s \in [n]$ be a node.

\begin{definition}[Flooding Process] \label{def::flo}
 The flooding process in $\mathcal{G}$ with source $s$ is the sequence $\{I_t \,:\, t \in \mathbb{N}\}$ of sets of nodes defined recursively as follows
\begin{itemize}
\item $I_0 = \{ s \}$ (at the beginning only the source node is \emph{informed});
\item $I_{t+1} = I_t \cup N_{t+1}(I_t)$ (the neighbours of informed nodes become informed)
\end{itemize}
where $N_{t+1}(I_t)$ is the neighborhood of $I_t$ in graph $G_{t+1}$,
$$
N_{t+1}(I_t) = \{ v \in [n] - I_t \;| \; \{u,v\} \in E_{t+1} \mbox{ for some } u \in I_t \}
$$
\end{definition}

\noindent Given an evolving graph $\mathcal{G}$ and a source node $s \in [n]$, the \emph{completion time} $T(\mathcal{G},s)$ of the flooding process is the first time step in which all nodes of the network are informed. The \emph{flooding time} $T(\mathcal{G})$ is the maximum completion time over all possible choices of source $s \in [n]$.

\subsection{Upper and lower bounds holding for any starting edge set}
 
In the following theorem we give upper and lower bounds on the flooding time of Home-MEGs, without any specific assumption on the initial set of edges $E_0$. Notice that this implies that the
obtained bounds  also holds  over the \emph{transient} phase of the Home-MEG.

\begin{theorem}\label{prop:ubflood}
Let $\mathcal{H}=\mathcal{H}(n,p,q,\alpha,\gamma)$ be a Home-MEG with $p+q \leqslant 1$ and $\gamma \leqslant \alpha$. Then, the flooding time $T_n(\mathcal{H})$ of $\mathcal{H}$ is 
$$
T_n(\mathcal{H}) = 
\left\{
\begin{array}{cl}
\mathcal{O}\left( \frac{\log n}{\log(1 + n \hat{p})} \right) &
\mbox{w.h.p.}\\[2mm]
\Omega\left( \frac{\log n}{\log(1 + n \hat{q})} \right) & \mbox{in expectation}
\end{array}
\right.
$$
where $\hat{p} = p \alpha + (1-p) \gamma$ and $\hat{q} = (1-q)\alpha + q\gamma$.\footnote{In the following, we use w.h.p. to denote the fact that
an event $E_n$, depending on a parameter $n$, holds with probability $\Prob{E_n} \geqslant  1- \frac{1}{n}$,  for every $n$ sufficiently large.}
\end{theorem}
\proof Recall that the random variable $X_t(e)$ describes the state of  edge $e \in \binom{[n]}{2}$ at time $t$, according to transition matrix (\ref{eq:EdgeHomeMEGtm}). Then, the probability edge $e$ exists at time $t$ is
$$
\Prob{X_t(e) \in \mathbf{C} \,|\, X_{t-1}(e) = x} =
\left\{
\begin{array}{cl}
(1-q) \alpha + q \gamma & \mbox{ if } x \in \mathbf{H} \\[2mm]
p \alpha + (1-p) \gamma & \mbox{ if } x \in \mathbf{N}
\end{array}
\right.,
$$
where $\mathbf{C}=\{HC, NC\}$, $\mathbf{H}=\{ HC, HD \}$, and $\mathbf{N}=\{ NC, ND \}$.
From the assumptions $p+q \leqslant 1$ and $\gamma \leqslant \alpha$, it follows that $\hat{q} = (1-q) \alpha + q \gamma \geqslant p \alpha + (1-p) \gamma = \hat{p}$. Hence, the probability an edge exists at time $t$ is at least $\hat{p}$ and at most $\hat{q}$, regardless of the state of the edge at time $t-1$.

\smallskip
Now consider two random evolving graphs $\mathcal{G}^p = \{ G_t^p \,:\, t \in \mathbb{N} \}$ and $\mathcal{G}^q = \{ G_t^q \,:\, t \in \mathbb{N} \}$ where $G_t^p$'s and $G_t^q$'s are independent Erd\"os-R\'enyi random graphs $G_{n,\hat{p}}$ and $G_{n,\hat{q}}$ respectively\footnote{An Erd\"os-R\'enyi $G_{n,p}$ random graph is a graph on $n$ nodes where, for every pair of nodes $u,v$ in the graph, undirected edge $(u,v)$ exists with probability $p$, independently of other edges.}. Since at every time step every edge exists in $\mathcal{H}$ with probability at least as large as in $\mathcal{G}^p$ and at most as large as in $\mathcal{G}^q$, it is intuitively true that the flooding time of $\mathcal{H}$ is at most as large as the flooding time of $\mathcal{G}^p$ and at least as large as the flooding time of $\mathcal{G}^q$. Since from~\cite{Clementi2} we know that the flooding time of $\mathcal{G}^p$ is $T_n(\mathcal{G}^p) = \mathcal{O} \left( \frac{\log n}{\log(1 + n \hat{p})} \right)$ w.h.p. and the flooding time of $\mathcal{G}^q$ is $T_n(\mathcal{G}^q) = \Omega \left( \frac{\log n}{\log(1 + n \hat{q})} \right)$ in expectation (see Theorems~3.7 and~5.3 in \cite{Clementi2}), the thesis follows.

A formal proof can be derived  by using the {\em coupling} technique (e.g. see~\cite{Lindvall}). Roughly speaking, one can define random processes $\mathcal{H}$, $\mathcal{G}^p$ and $\mathcal{G}^q$ on the same probability space, in a way that, naming $E_t(\mathcal{H})$, $E_t(\mathcal{G}^p)$, and $E_t(\mathcal{G}^q)$ the (random) sets of existing edges at time $t$ in processes $\mathcal{H}$, $\mathcal{G}^p$ and $\mathcal{G}^q$ respectively, it holds that $E_t(\mathcal{G}^p) \subseteq E_t(\mathcal{H}) \subseteq E_t(\mathcal{G}^q)$ for every $t$.  In what follows we make the above argument rigorous.
 
 We need some technical machinery, namely the \emph{coupling} technique (e.g. see~\cite{Lindvall}). Roughly speaking, we will define random processes $\mathcal{H}$ and $\mathcal{G}$ on the same probability space, in a way that, if we name $E_t^{\mathcal{H}}$ and $E_t^{\mathcal{G}}$ the (random) sets of existing edges at time $t$ in processes $\mathcal{H}$ and $\mathcal{G}$ respectively, it will hold that $E_t^{\mathcal{G}} \subseteq E_t^{\mathcal{H}}$ for every $t$. That can be achieved, for example, as follows.

 \begin{figure}[h]
 \begin{center}
 \includegraphics[width=12cm]{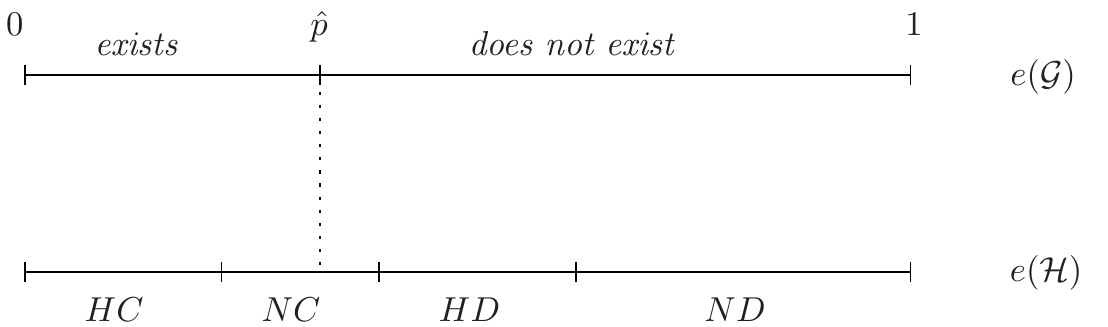}
 \end{center}
 \caption{Coupling for the edge $e$ at time $t$ for processes $\mathcal{H}$ and $\mathcal{G}$}\label{fig:coupling}
 \end{figure}

\noindent Let $\{U_t(e)\,:\, t \in \mathbb{N}, \, e \in \binom{[n]}{2} \}$ be a family of independent random variables uniformly distributed over the interval $[0,1]$; at time $t$ we use random variable $U_t(e)$ to update the state of edge $e$ in both processes $\mathcal{H}$ and $\mathcal{G}$ as illustrated in Figure~\ref{fig:coupling}:
\begin{enumerate}
\item We take two copies of interval $[0,1]$ and we name them $e(\mathcal{H})$ and $e(\mathcal{G})$;
\item We partition the $e(\mathcal{G})$-interval in two sub-intervals: the leftmost one, of length $\hat{p}$, labeled \emph{exists} and the rightmost one, of length $1-\hat{p}$ labeled \emph{does not exist};
\item We partition the $e(\mathcal{H})$-interval in four sub-intervals: the two on the left labeled $HC$ and $NC$ and the two on the right labeled $HD$ and $ND$. The lengths of the sub-intervals are chosen according to the transition probabilities from the previous state of edge $e$ in process $\mathcal{H}$. For example, if edge $e$ at time $t-1$ was in state $HC$, then the lengths are the probabilities in the first row of the transition matrix (\ref{eq:EdgeHomeMEGtm});
\item We set the state of edge $e$ at time $t$ in processes $\mathcal{H}$ and $\mathcal{G}$ according to the label of the sub-interval containing $U_t(e)$ in intervals $e(\mathcal{H})$ and $e(\mathcal{G})$ respectively.
\end{enumerate}
By construction, the probability edge $e$ exists in process $\mathcal{G}$ is $\hat{p}$, and the probability edge $e$ is in one of the four states in process $\mathcal{H}$ agrees with the transition matrix (\ref{eq:EdgeHomeMEGtm}). Moreover, since the sum of the lengths of sub-intervals $HC$ and $NC$ is always at least as large as $\hat{p}$, if the edge exists in process $\mathcal{G}$ then it exists in process $\mathcal{H}$; hence, at every time step $t$ we have that $E_t^{\mathcal{G}} \subseteq E_t^{\mathcal{H}}$ . Thus, if the flooding process in $\mathcal{G}$ is completed within time $\bar{t}$, then the flooding in $\mathcal{H}$ has to be completed within time $\bar{t}$ as well.
\qed

\subsection{A tighter upper bound for the stationary case (and slow changes)}\label{sec:slow}
In the previous subsection, we have provided upper and lower bounds on  flooding time for the Home-MEG model. Notice that, while the upper bound stated in Theorem~\ref{prop:ubflood} is not based on any specific assumption about the initial edge set $E_0$, it is not tight in general: when $p+q \ll 1$ and $\gamma < \alpha$ -- which is the case in practical situations, recall the discussion at the end of the previous section -- then $p \alpha + (1-p) \gamma \ll (1-q) \alpha + q \gamma$ and the lower bound on the probability of the existence of an edge we used in Theorem~\ref{prop:ubflood} is too lose.
Indeed, consider the following example: assume $p =  q =  \alpha = 1/n$ and $\gamma = 1/n^2$. Notice that $p = q$ means that, if $E_0$ is random according to the stationary distribution, edges are in Home location approximately half of the time. In this case we have
$$
\hat{p} = p\alpha + (1-p)\gamma = \gamma + p (\alpha - \gamma) \approx \frac{1}{n^2}
$$
and Theorem~\ref{prop:ubflood} gives an upper bound $\mathcal{O}(n\log n)$. On the other hand, the probability of existence of an edge, if it was in Home location at previous time step, is
$$
(1-q) \alpha + q \gamma = \alpha - q (\alpha - \gamma) \approx \frac{1}{n}
$$
Hence, based on \cite{Clementi2}, one would expect flooding time to be something like $\mathcal{O}(\log n)$ in this case.
In other words, in the proof of Theorem~\ref{prop:ubflood} we lower bounded the probability of existence of an edge in a given time step as the edge always was in Non-Home state at previous time step, and this is clearly a loose bound when the ``stationary'' expected fraction of   edges   in Home state is significant and $\gamma \ll \alpha$. In order to obtain a tighter upper bound on the flooding time, we instead fully exploit the properties of the stationary configuration and get the bound stated in the theorem below. For convenience sake, in what follows we set 
\[ \Lambda = \frac{4(p+q)}{p \alpha}\] Observe that, according to~(\ref{eq:lbprob}), we have $4/\Lambda = \pi(HC)$.

\begin{theorem}\label{thm::flotime}
Let $\mathcal{H}=\mathcal{H}(n,p,q,\alpha,\gamma)$ be a Home-MEG and 
assume that $ \lceil    \frac {5 \Lambda }{   n  } \rceil \leqslant  \min\{ 1/\alpha , 1/(4q) \}$.
Then, if the initial edge set $E_0$ is random according to the stationary distribution $\pi$, the flooding  time $T_n(\mathcal{H})$ is w.h.p.  
\begin{equation*} 
T_n(\mathcal{H}) =  \mathcal{O}\left( \frac{\log n}{\log (1 + n/\Lambda)} \right)
\end{equation*}
\end{theorem}

 The upper bound in Theorem \ref{thm::flotime} is not always applicable (see the condition on $\Lambda$ and on $E_0$) and,
in   case of frequent changes between the Home and the Non-Home states (i.e. for large values of $p$ and $q$),
it can be much  weaker than the upper bound of Theorem \ref{prop:ubflood}. For instance, for $\gamma \sim 1/n$, $\alpha \sim 1/n^{2/3}$,
$p\sim 1/n^{4/3}$, and $q \sim1/n^{2/3}$, the bound of   Theorem \ref{thm::flotime} is $\Theta(n^{1/3}\log n)$ while the upper bound
obtained by Theorem \ref{prop:ubflood} is $\mathcal{O}(\log n)$.

On the other hand, the upper bound in Theorem  \ref{thm::flotime} can be much  tighter than that of Theorem~\ref{prop:ubflood} in the realistic scenarios described in 
Section \ref{sec:fitting}. Indeed,
 we  first  observe that the assumption $ \lceil    \frac {5 \Lambda }{   n  } \rceil \leqslant  \min\{ 1/\alpha , 1/(4q) \}$
is in accordance with the ``realistic'' assumptions on $p,q,\alpha,\gamma$ derived in Section \ref{sec:fitting}:  When $p+q \ll 1$, the assumption is satisfied for any $p \geqslant q / n$.
More importantly,  the above theorem has the following interesting consequence.

\begin{cor}\label{HMEGBound}
Consider an opportunistic network with $n$ nodes, and assume pairwise link occurrences are modeled with the Home-MEG model, with parameters set as follows:
 \[\alpha =\frac{ n^{\epsilon}}{n}, \ \gamma =\frac{1}{n^2}, \
p=\frac{1}{ n^{1+\epsilon}}, \ q=\frac{1}{n} \] for some arbitrary constant $0<\epsilon <1$. Then, flooding in the network is w.h.p. completed in $\mathcal{O}(\log n)$ time.
\end{cor}

\smallskip
We observe  that all conditions $i)$, $ii)$, and $iii)$ stated in Subsection \ref{subse:val}
are met when the parameters of the Home-MEG model are set as in the statement of Corollary~\ref{HMEGBound}. Also, when the parameters of the Home-MEG model are set as above, the stationary expected edge  density (i.e. the number of neighbors of a node) is $1$, i.e., the network is well below the connectivity threshold of $\mathcal{O}(\log n)$ (this is  consistent with the opportunistic network setting). With similar parameter settings, Theorem \ref{prop:ubflood} gives a bound on flooding time of $\mathcal{O}(n\log n)$. 

Our bound on flooding time  for sparse Home-MEG  leads to conclusions that largely depart from those arising from simple geometry mobility models. Indeed, in \cite{Clementi,Clementi1,PSSS10,PPPU11}, tight  bounds on flooding time are derived based on the random walk model in which the mobility and transmission range of nodes can be fixed in a certain range. Informally speaking, $n$ nodes move in a square region of side $\sqrt{n}$, nodes transmission range is $R$, with $c\le R \le \sqrt{n}$, and the nodes mobility radius in one time step is $\mathcal{O}(R)$. 
Under these conditions, the flooding time is shown  to be $\Theta\left(\frac{\sqrt{n}}{R}\right)$. By setting the transmission range in such a way that the expected node degree is $\Theta(1)$  (i.e. the same as in Corollary \ref{HMEGBound})  the results of \cite{Jacquet,PSSS10,PPPU11} yield a $\Theta(\sqrt{n})$ bound on flooding time, which is exponentially larger  than the   bound obtained ($O(\log n)$ vs $O(\sqrt{n})$) with the Home-MEG model (Corollary \ref{HMEGBound}).
It is important to observe that both existing bounds based on geometric mobility models and the bounds derived herein display both realistic as well as unrealistic features of the induced contact patterns between nodes, as already discussed in Section \ref{sec:fitting}. In particular, geometric mobility models faithfully capture the fact that occurrence/disappearance of a link between a pair of nodes is governed by physical notions such as distance between nodes, transmission range, etc. On the other hand, the specific mobility models used in the existing flooding time analyses \cite{Clementi,Clementi1,PSSS10,PPPU11} are known to be unrealistic mainly because of two reasons: 
(i) the generated contact traces do not display the power law+exponential tail dichotomy of inter-contact times, and (ii) human and veichular mobility largely differs from basic random walks. The Home-MEG model used in our analysis tackles the issue of the inter-contact time dichotomy, but it is an abstract contact model in which occurrence of links between nodes is not related with physical notions such as distance between nodes, transmission range, etc., thus it is not realistic in that respect. Despite this unrealistic aspect, the experimental results presented in Section \ref{sec:fitting} indicate that the Home-MEG model is a reasonable candidate to estimate flooding time in opportunistic networks, while theoretical analysis of this model suggests that flooding time in opportunistic networks might be much shorter than predicted by existing polynomial bounds obtained from simple geometric mobility models. 

We want to comment on the fact that the Home-MEG model parameters are assumed to be dependent on the number $n$ of nodes in the network, and in particular decreasing with $n$. This assumption appears reasonable, since a single Markov chain in Home-MEG models the existence/non-existence of the link between a specific pair of nodes in the network. For instance, this assumption is in accordance with a scenario in which the number of nodes (each with a fixed transmission range) in the network grows, while the density of nodes per unit area remains unchanged (constant density networks).
Finally, we emphasize that, differently from previous related work~\cite{Jacquet,PSSS10}, our theoretical analysis provides  high concentration results on flooding time (the results hold \emph{with high probability} rather than \emph{asymptotically almost surely}\footnote{An event $E_n$ holds \emph{a.a.s.} if $\lim_{n\to\infty}\Prob{E_n}=1$.})  and the constants hidden by the asymptotic notation 
are rather small, say not larger than 10. Both these properties are clearly crucial when the goal is to apply  theoretical results to realistic scenarios.

\subsubsection{Proof of Theorem \ref{thm::flotime}}
We first  provide  an informal description  of the main arguments.

\noindent
Let $\mathcal{H}=\mathcal{H}(n,p,q,\alpha,\gamma)$ be a Home-MEG  and, for any    edge  $e$ and positive integer $\ell$, define the ``connection'' probability $P_{\ell}(e) = \Prob{\exists i \in \{ 0, \ldots , \ell \} \ | \ e \in E_i  }$. By assuming $E_0$ to be random with the stationary distribution, we prove that
\[
P_{\ell}(e) \geqslant  \frac{p \alpha}{p+q} \ell \ \mbox{ for any } \ \ell \leqslant   \min\{ 1/\alpha , 1(4q) \}
\]
The above bound says that, starting from the stationary distribution, the connection probability grows  linearly in $\ell$, for a while. A natural approach would be that of exploiting this bound to derive some good node-expansion property of the time evolving graph and, then, applying this expansion to the growing subset of informed nodes along the flooding process.
This approach implies that, starting from the stationary distribution, at any time of the flooding process, we are interested in the edges that connect informed to uninformed nodes. This in turn entails that the state of the same edge might be observed multiple times. Unfortunately, when the state of an edge is observed at a given time, then it cannot be considered ``stationary'' anymore. We overcome this technical difficulty by considering a ``subprocess'' of the flooding which is organized in \emph{time-periods}: at the beginning of every time-period, only ``fresh'' edges are considered, i.e., edges that have been never observed before. More precisely, let $C_i$ be the set of nodes that have been informed during the $i$-th time-period, then only the edges in the cut  $(C_t, V - \cup_{i=0}^{t-1} C_i)$ will be considered in time period $t$. Finally, by setting the length of the time-periods in a suitable way, we get the bound.
   
\medskip\noindent In the remainder of this section we provide the formal proof. We will make use of the following well-known inequalities: For any positive real  $0<x<1$, it holds that
\begin{equation}\label{eq::usef1}
1-x \leqslant e^{-x} \leqslant  1 - x/2
 \end{equation}
We first provide a general  upper bound on the edge-disconnection time  for any  finite  Markov chain governing the edge evolution.
Let $A$ be any subset of the states of the Markov chain and let $\mathbf{D}$ be the subset of states implying edge disconnection.
Notice that in the Home-MEG model, $\mathbf{D} = \{ HD,ND\}$. Assume the starting edge configuration $E_0$ is random with the stationary distribution $\pi$ and, for any state subset $X$ of the Markov chain, define  
\begin{equation*}
\sE_{\ell}^X = ``e \in X , \   \forall t = 0, \ldots , \ell ''      
\ \mbox{ and }  \   \pi^{A,\mathbf{D}}_{\ell} \ = \ \Prob{  \sE_{\ell}^\mathbf{D} \ | \ \ \sE_0^A   }
\end{equation*}

\begin{lemma} \label{lem::disctime}
Let $A$ and $\mathbf{D}$ be such that:  
\begin{itemize}
\item Hyp. $I.$ $\ \forall s \in A \cap \mathbf{D}, \ \   \Prob{s \rightarrow A \cap \mathbf{D}} \leqslant \lambda $;
\item Hyp $II.$ $\ \forall s \in A, \ \   \Prob{s \rightarrow A} \geqslant \delta $. 
\end{itemize}
Then, for any $\ell \geqslant 1$ it holds that
\begin{equation}\label{eq::disctime}
\pi^{A,\mathbf{D}}_{\ell} \leqslant \ 1 - \delta^{\ell} \left( 1- \frac{\lambda^{\ell}}{\delta^{\ell}} \frac{\pi(A \cap\mathbf{D})}{\pi(A)}  \right)
\end{equation}
\end{lemma}
\proof It holds that
\begin{eqnarray*} \label{eq::p1}
\pi^{A,D}_{\ell} & = & \Prob{ \sE_{\ell}^D | \sE_{0}^A \wedge  \sE^A_{\ell}} \Prob{\sE^{A}_{\ell} | \sE_{0}^A } + \Prob{ \sE_{\ell}^D  | \sE_{0}^A \wedge \neg \sE^A_{\ell}} \Prob{\neg \sE^{A}_{\ell} | \sE_{0}^A} \\[2mm]
& \leqslant & \Prob{ \sE_{\ell}^D \ | \ \sE_{0}^A \wedge  \sE^A_{\ell} } \Prob{\sE^{A}_{\ell} |  \sE^A_{0}} + 1 \cdot  (1 -  \Prob{\sE^A_{\ell} |  \sE^A_{0} })  
\end{eqnarray*}
So,
\begin{equation} \label{eq::B1}
\pi^{A,D}_{\ell} \ \leqslant \ 1 - \Prob{\sE^A_{\ell} |  \sE^A_{0} } (1 -\Prob{\sE^D_{\ell} | \sE^A_{\ell}})
\end{equation}
Let us first bound $\Prob{\sE^A_{\ell}}$.
\begin{eqnarray*}
\Prob{ \sE^A_{\ell} } &  =  &  \Prob{ \sE^A_{\ell} | \sE^A_{0}   \wedge  \sE^A_{\ell-1} } \Prob{  \sE^A_{\ell-1} | \sE^A_{0} } + \Prob{ \sE^A_{\ell} | \sE^A_{0}   \wedge  \neg \sE^A_{\ell-1}  } \Prob{ \neg \sE^A_{\ell-1}| \sE^A_{0} }  \\[2mm]
& = & \Prob{ \sE^A_{\ell} |   \sE^A_{\ell-1} }  \Prob{ \sE^A_{\ell-1} | \sE^A_{0} } \\
\mbox{ (by Hyp. II) } & \geqslant & \delta  \ \Prob{ \sE^A_{\ell-1} |\sE^A_{0} }
\end{eqnarray*}
So, we get
\begin{equation} \label{eq::L1}
\Prob{\sE^A_{\ell}} \ \geqslant \ \delta^{\ell}
\end{equation}
As for  the term $\Prob{\sE^D_{\ell} | \sE^A_{\ell}}$, it holds
\begin{equation}\label{eq::Ub1}
\Prob{ \sE^D_{\ell} | \sE^A_{\ell} } = \frac{ \Prob{ \sE^D_{\ell} \wedge \sE^A_{\ell}  }} {  \Prob{\sE^A_{\ell}}} =
\frac{ \Prob{  \sE^{D \cap A}_{\ell }} }{ \Prob{ \sE^A_{\ell} } }
\end{equation}
while, for $\Prob{  \sE^{ D \cap A}_{\ell} } $, we apply Hyp. I and get
\begin{eqnarray*}
\Prob{  \sE^{D\cap A}_{\ell} } & = &  \Prob{\sE^{ D\cap A}_{\ell} | \sE^{D\cap A}_{0} } \Prob{\sE^{D\cap A}_{0}}
+ \Prob{\sE^{ D\cap A}_{\ell}  | \neg \sE^{D\cap A}_{0}} \Prob{ \neg \sE^{D\cap A}_{0} } \\[2mm]
& =  &   \Prob{ \sE^{ D\cap A}_{\ell} |\sE^{ D\cap A}_{\ell-1} } \cdot \Prob{\sE^{ D \cap A}_{\ell-1} | \sE^{D\cap A}_{0} } \Prob{\sE^{D\cap A}_{0}} \\[2mm]
\mbox{ (by Hyp I) } & \leqslant  & \lambda  \  \Prob{ \sE^{ D \cap A}_{\ell-1} | \sE^{D\cap A}_{0} }\pi(A \cap D)  
\end{eqnarray*}
We thus get 
\begin{equation}\label{eq::Ub2}
\Prob{  \sE^{D \cap A}_{\ell} } \leqslant \lambda^{\ell}  \pi(A \cap D)  
\end{equation}
As for $\Prob{\sE^A_{\ell} }$, 
\begin{eqnarray*}
\Prob{ \sE^A_{\ell} } & = &  \Prob{ \sE^A_{\ell} | \sE^A_{0} } \Prob{\sE^A_{0}} + 0 \\
\mbox{From Eq. \ref{eq::L1}} & \geqslant & \delta^{\ell} \pi(A)
\end{eqnarray*}
By combining the above bound with Ineq. \ref{eq::Ub2}, Ineq.  \ref{eq::Ub1} becomes
\[
\Prob{ \sE^D_{\ell} | \sE^A_{\ell} } \leqslant  \frac{\lambda^{\ell}}{\delta^{\ell}} \frac{\pi(A \cap D)}{\pi(A)}
\]
Finally, by replacing the above bound and that in \ref{eq::L1} in Ineq \ref{eq::B1},
we obtain
\[
\pi^{A,D}_{\ell} \leqslant \ 1 - \delta^{\ell} \left( 1- \frac{\lambda^{\ell}}{\delta^{\ell}} \frac{\pi(A \cap D)}{\pi(A)}  \right)
\]
\qed

\noindent
The above lemma can be applied to a Home-MEG   
$\mathcal{H}=\mathcal{H}(n,p,q,\alpha,\gamma)$ with  $A = \mathbf{H} =  \{ HC,HD\}$ and $\mathbf{D}$ as above. Notice that
\begin{eqnarray}\label{eq::D-H}
\pi(\mathbf{H})  =  \frac{p}{p+q} & \mbox{and} &  \pi (\mathbf{H} \cap \mathbf{D})   =  \frac{p(1-\alpha)}{p+q}  \\
\lambda =   (1-q)(1-\alpha) & \mbox{and} & \delta   = (1-q)
\end{eqnarray}
   
\noindent
By replacing the above formulas in Ineq. (\ref{eq::disctime}), we get the following
\begin{cor}
Let $\mathcal{H}=\mathcal{H}(n,p,q,\alpha,\gamma)$ be a Home-MEG. Then, for any $\ell \geqslant 1$
it holds 
\begin{equation} \label{eq::dtimehome}
\pi^{\mathbf{H},\mathbf{D}}_{\ell} \leqslant 1 - (1-q)^{\ell}(1 - (1-\alpha)^{\ell})
\end{equation}
\end{cor}

\paragraph{The Flooding Process.}
In order to manage stochastic dependence, we consider a somewhat \emph{sub-process} of the actual
flooding by (only) looking at restricted subsets of informed nodes and edges. This subprocess works in consecutive \emph{time periods}  $\Delta_{\tau}$, $\tau = 0, 1, \ldots$: each time period $\Delta_\tau$ is  formed by the time steps $t_\tau,  t_{\tau} +1, \ldots , t_\tau + L_\tau-1$.  
The period lengths $L_\tau$ are defined as follows. Define
\[
K = 2 \max \left\{1, \frac{ n}{5 \Lambda } \right \}, \ 
t_0 = 0,  \ t_\tau =  t_{\tau -1} + L_{\tau - 1 }
\]
where  $L_1 =   \left \lceil  \frac  { 4 \Lambda    \log n}{n} \right \rceil$ and, for $\tau \geqslant 2$
\begin{equation} \label{eq::time}
L_\tau \ = \ 
\left\{
\begin{array}{cl}
1 & \quad \mbox{ if }\,  2 K^{\tau -2}  \log n \geqslant \Lambda \\
\left\lceil \frac {5 \Lambda}{n} \right\rceil & \quad \mbox{ otherwise}
\end{array} \right .
\end{equation}

\noindent
Then, for any time period $\Delta_\tau$, define the following node subsets 

\begin{eqnarray*}
C_0 & = & \{s \} \\[2mm]
C'_{\tau} & = & \{ v \in V \setminus \cup_{0}^{\tau -1}C_{i} \ | \exists t \in \Delta_\tau \mbox{ and  } w  \in C_{\tau -1} \ \mbox{ s.t. } (w,v) \in E_{t-1} \} \\[3mm]
C_{\tau} & = &
\left\{  
\begin{array}{ll}
C'_{\tau} & \mbox{if }   |C'_{\tau}| \leqslant   2 K^{\tau -1}  \log n \ \mbox{ or } \  |C'_{\tau}| \geqslant n/16  \\[2mm]
\hat{C}'_{\tau}   &   \mbox{otherwise}    
\end{array}
\right .
\end{eqnarray*}
where $\hat{C}'_\tau$ is any subset of  $C'_\tau$ of size $2 K^{\tau -1}  \log n $.
  
\noindent
Observe that subsets $C_{\tau}$ are mutually disjoint and the  edges in the cut  $E(C_{\tau - 1}, C_\tau)$ are not considered by the (sub-)flooding process before time-period $\tau$, so in that period they are random with the stationary distribution. Moreover, $C_\tau$ is a subset of the set of informed nodes at the end of time period $\tau$.

\smallskip\noindent
\textbf{Phase 1.}
We now analyze the size of $C_1$. This time period  of the flooding process will be called \emph{Bootstrap}.

\noindent
Observe that, by definition of $ \pi^{H,D}_{\ell}$, Ineq.~\ref{eq::dtimehome} holds under the condition
$e \in H$ at time $t=0$. So,
\[\Prob{\sE^D_{\ell} }  =  \Prob{\sE^D_{\ell} | \sE^H_{0} } \Prob{\sE^H_{0} } + \Prob{\sE^D_{\ell} | \neg \sE^H_{0} } \Prob{\neg \sE^H_{0} } \]
\begin{eqnarray*}
\mbox{ (by Ineq. \ref{eq::dtimehome}) } \hspace{-2mm}& \leqslant &  \pi^{H,D}_{\ell} \pi(H) + (1-\pi(H))
\end{eqnarray*}

\noindent
By applying Ineq. \ref{eq::dtimehome} and Eq. \ref{eq::D-H}, we get 
\begin{equation} \label{eq::dtimeh2}
\Prob{\sE^D_{\ell} } \  \leqslant \ 1 -\frac{p}{p+q}(1-q)^{\ell}(1-(1-\alpha)^\ell) 
\end{equation}
 
\noindent
We now assume that $\ell \leqslant \min\{ 1/\alpha , 1(4q) \}$. 
Then, we apply Ineq. \ref{eq::usef1} to Ineq. \ref{eq::dtimeh2} and get the following  

\begin{lemma} \label{lm::dtimeapx}
Let  $\ell \leqslant \min\{ 1/\alpha , 1/(4q) \}$, then it holds that
 \begin{equation} \label{eq::dtime-apx}
\Prob{\sE^D_{\ell} } \  \leqslant \ 1 - \frac{\ell}{ \Lambda }  
\end{equation}
\end{lemma}

\noindent
By using the above lemma, we  can prove an upper bound on the bootstrap time.

\begin{lemma}\label{lm::boot}
Let $\mathcal{H}=\mathcal{H}(n,p,q,\alpha,\gamma)$ be a Home-MEG.
Then   it holds that  
\begin{equation*}
\Prob{|C_1|  \leqslant 2 \log n }  \ \leqslant \ \frac {1}{n^2}
\end{equation*}
Thus, the time of Phase 1 is w.h.p. upper bounded by $L_1 =   \left \lceil  \frac  { 4 \Lambda    \log n}{n} \right \rceil$.
\end{lemma}
\proof
Observe that $\ell = L_1$ satisfies the hypothesis of     Lemma \ref{lm::dtimeapx}.
We thus apply  this lemma for $\ell = L_1$  and get that the  $\Prob{\sE^D_{L_1} } \leqslant
1-\frac{ 4 \log n}{n}$.

\noindent
Let us now consider every possible edge of the source node and look at its behaviour for the first $L_1$ time steps.
Since edges  are mutually independent, we can apply Chernoff's bound \cite{MU05} and, thanks to Ineq. \ref{eq::dtime-apx}, we get that, w.h.p., the number of nodes that will be informed by the source node within the first $L_1$ steps is at least $2 \log n$.
\qed

\smallskip \noindent
\textbf{Phase 2.}  We can now assume there are at least  $2 \log n$ informed nodes and we analyze the flooding process till an arbitrary constant fraction of all the $n$ nodes are informed. This is the second phase.
 
\noindent
For every $\tau \geqslant 1$, define the events
$$
\begin{array}{cclccl}
F_\tau & = & \virgl |C_\tau | \geqslant \frac{n}{16} \virgr \, & X_\tau & = & \virgl |C_\tau | \geqslant 2 K^{\tau -1 }  \log n  \virgr \\[2mm]
\widehat{F}_\tau & = &  \bigvee_{t = 0}^\tau F_t \,  & \widehat{X}_\tau & = & \bigwedge_{t = 0}^{\tau} X_t
\end{array}
$$
where $X_0$ is the event $\virgl |C_0| \geqslant 0\virgr$ (which happens with probability $1$).

\begin{lemma}\label{lem::expansion}
Assume that $ \lceil    \frac {5 \Lambda }{   n  } \rceil \leqslant  \min\{ 1/\alpha , 1/(4q) \}$.
For every $\tau \geqslant 1$, it holds that
$$
\Prob{(F_\tau \vee X_\tau) \wedge \neg \widehat{F}_{\tau - 1} \wedge \widehat{X}_{\tau - 1}}  \geqslant \left(1 - \frac{1}{n^2}\right) \Prob{\neg \widehat{F}_{\tau - 1} \wedge \widehat{X}_{\tau - 1}}
$$
\end{lemma}
\proof
Define the event
\[
B = \virgl \sum_{t = 0}^{\tau - 1} |C_t| < \frac{n}{8}\virgr
\]
We show that the following implication holds
\begin{equation}\label{implication}
\neg B \wedge  \widehat{X}_{\tau - 1} \quad \Rightarrow \quad  \widehat{F}_{\tau - 1}
\end{equation}
If $\tau = 1$, it trivially holds since $ \neg B$ is false. 
As for  $\tau \geqslant 2$, let us assume by contradiction that event $\widehat{F}_{\tau-1}$ does not hold.
Since  $ \widehat{X}_{\tau - 1}$ holds, according to the definition of $C_\tau$,  it must be the case  
$|C_{\tau}| =  K |C_{\tau -1 }| $ with 
$K \geqslant 2$. This implies   that 
\[
|C_{\tau - 1}| \geqslant \sum_{t = 0}^{\tau - 2} |C_t|
\]
It follows that 
\[
|C_{\tau - 1}| \geqslant \frac{1}{2}( |C_{\tau - 1}| + \sum_{t = 0}^{\tau - 2} |C_t|) = \frac{1}{2}\sum_{t = 0}^{\tau - 1} |C_t|
\]
Since $ \neg B$ holds, we have that
\[
|C_{\tau - 1}| \geqslant \frac{1}{2}\sum_{t = 0}^{\tau - 1} |C_t| \geqslant \frac{n}{16}
\]
and $\widehat{F}_{\tau - 1}$ holds, a contradiction.
 
\noindent It is immediate to see that (\ref{implication}) implies the following
\begin{equation*}
\neg \widehat{F}_{\tau - 1} \wedge \widehat{X}_{\tau - 1} \quad \Rightarrow \quad B
\end{equation*}
It derives that
$$
\Prob{(F_\tau \vee X_\tau) \wedge \neg \widehat{F}_{\tau - 1} \wedge \widehat{X}_{\tau - 1}} = \Prob{(F_\tau \vee X_\tau) \wedge \neg \widehat{F}_{\tau - 1} \wedge \widehat{X}_{\tau - 1}\wedge B}
$$
and
\begin{equation*}
\Prob{\neg \widehat{F}_{\tau - 1} \wedge \widehat{X}_{\tau - 1}} =
\Prob{\neg \widehat{F}_{\tau - 1} \wedge \widehat{X}_{\tau - 1} \wedge B}
\end{equation*}
Moreover, if 
\[
\Prob{\neg \widehat{F}_{\tau - 1} \wedge \widehat{X}_{\tau - 1}\wedge B} = 0
\]
then the thesis trivially holds. Thus, suppose that
\[
\Prob{\neg \widehat{F}_{\tau - 1} \wedge \widehat{X}_{\tau - 1}\wedge B} > 0 
\]
Then, define event 
$\sW = \neg \widehat{F}_{\tau - 1} \wedge \widehat{X}_{\tau - 1} \wedge B$, then  the thesis can be reformulated as follows
\begin{equation*}
\Prob{F_\tau \vee X_\tau\; \left|\;  \sW \right.} \geqslant 1 - \frac{1}{n^2}
\end{equation*}

\noindent
We first consider the case $\tau = 1$.  In this case, we get 
  
\begin{eqnarray*} 
\Prob{F_1 \vee X_1\; \left|\;  \sW \right.} & =&  \\
\mbox{ since $\Prob{W} =1$  } & \geqslant &  \Prob{F_1 \vee X_1} \geqslant \Prob{|C_1| \geqslant 2 \log n } \\ 
\mbox{ from Lemma \ref{lm::boot}}   & \geqslant &  1 - \frac{1}{n^2}
\end{eqnarray*}
  
\noindent
Consider now a time period $\tau \geqslant 2$ and, for any node $v \in V \setminus \cup_{0}^{\tau -1}C_{i}$,   define r.v. $Y_v=1$ iff $v \in C'_\tau$.
 
\noindent
Since $ \lceil    \frac {5 \Lambda }{   n  } \rceil \leqslant  \min\{ 1/\alpha , 1/(4q) \}$, $L_\tau$ satisfies the hypothesis of Lemma~\ref{lm::dtimeapx}. It holds that
       
\begin{eqnarray}\label{eq::expecII}
\Expec{\left.  |C'_\tau |  \; \right  |  \  \sW   } & =  &   \Expec{\left . \sum_v Y_v \right | \ \sW}  \nonumber \\
& = & \Expec{ \left . \sum_{ v \in V \setminus \cup_{0}^{\tau -1}C'_{i} \  |} \left( 1 - \left(1- \frac{L_\tau}{ \Lambda}  \right)^{|C'_{\tau -1}|} \right) \right  |    \ W} \nonumber \\
& \geqslant & \hspace{5mm} \Expec{ \left . \frac {7n}8 \left(1 - \exp{\left( - \frac{  L_\tau  |C'_{\tau -1}| }{\Lambda}\right)  }\right) \ \right | W }
\end{eqnarray}

\noindent
Two cases may arise.

\smallskip\noindent
{\bf Case A: $2 K^{\tau - 2}   \log n  \  \geqslant \  \Lambda$}. 
In this case $L_\tau =1$ and from    Ineq.  \ref{eq::expecII} we get
\begin{equation*}
\Expec{|C'_\tau|\  |  \ \sW }  \geqslant \ \frac {7 n}8 \left(1 -   e^{- L_{\tau}}\right)  
\  \geqslant \ \frac {7 n}8 \left(1 -  \frac 1e\right) \geqslant \frac {7 n}{16}
\end{equation*}

\smallskip\noindent
{\bf Case B:  $2 K^{\tau - 2}   \log n  \  <  \  \Lambda$.}
From Eq.~\ref{eq::time}, it holds 
\begin{equation*}
L_\tau  \ = \  \left\lceil  \frac{5 \Lambda} {  n   }  \right\rceil
\end{equation*}
From Ineq. \ref{eq::expecII}, we get

\[
\Expec{|C'_\tau| \ | \ \sW}   \geqslant   \Expec{ \left . 
\frac {7n}8 \left(1 - \exp{\left( - \frac{  \left\lceil  \frac{5 \Lambda} {  n   }  \right\rceil  |C'_{\tau -1}| }{\Lambda}\right)  }\right)  \ \right |  \ \sW}
\]

\noindent Observe that event  $\sW$ implies $\widehat X_{\tau -1}$ which in turn implies $|C'_{\tau - 1 }| \geqslant 2 K^{\tau -2}   \log n$.
It holds that
\begin{equation}\label{eq::CW}
\Expec{|C'_\tau| \ | \ \sW}   \geqslant      
\frac {7n}8 \left(1 - \exp{\left( - \frac{  \left\lceil  \frac{5 \Lambda} {  n   }  \right\rceil   2 K^{\tau -2}   \log n }{\Lambda}\right)  }\right)
\end{equation}
We again distinguish two subcases.  If $\Lambda \geqslant n/5$, 
from  the above inequality,  we get 
\[
\Expec{|C'_\tau| \ | \ \sW}   \geqslant 
\frac {7n}8 \left(1 - \exp{\left( - \frac{  10  K^{\tau -2}   \log n }{n}\right)  }\right)
\]
Since $\neg \widehat F$ implies  $	\frac{ 10  K^{\tau -2}   \log n }{n} \leqslant 1$, we can apply~(\ref{eq::usef1}) and get 
\[
\Expec{|C'_\tau| \ | \ \sW}   \geqslant 
\frac {7n}8    \frac{  10  K^{\tau -2}   \log n }{2n} = \frac{ 70    K^{\tau -2}   \log n }{ 16 } 	\geqslant \frac{70}{32} K^{\tau -1 }   \log n
\]
where the last inequality follows from the fact that, in this subcase, $K =2$.

\noindent
As for the second subcase   $\Lambda < n/5$, since $\lceil (5\Lambda)/n \rceil =1$, Ineq.~\ref{eq::CW} becomes
\[
\Expec{|C'_\tau| \ | \ \sW}   \geqslant      
\frac {7n}8 \left(1 - \exp{\left( - \frac{   2 K^{\tau -2}   \log n }{\Lambda}\right)  }\right)
\]
Since again $\neg \widehat F$ implies  $\frac{2 K^{\tau -2} \log n}{n} \leqslant 1$, we can apply~(\ref{eq::usef1}) and get 
\[
\Expec{|C'_\tau| \ | \ \sW} \geqslant 
\frac {35 n}8 \frac{2 K^{\tau -2} \log n}{5 \Lambda} = \frac{70 }{16} K^{\tau -1} \log n
\]

\noindent
In both cases, we can apply Chernoff's bound on the sum $C'_\tau$ of independent r.v.  and then,
thanks  to the definition of $C_\tau$,  get the thesis.  
\qed

\begin{lemma}\label{lem::expansion2}
Assume that $ \lceil    \frac {5 \Lambda }{   n  } \rceil \leqslant  \min\{ 1/\alpha , 1/(4q) \}$. Then,
for every $\tau \geqslant 1$, it holds that
\[ \Prob{\widehat{F}_{\tau} \vee \widehat{X}_{\tau}} \geqslant 1 - \frac{\tau}{n^2}  \]
\end{lemma}
\proof
For the sake of convenience, let $A_\tau = \widehat{F}_{\tau} \vee \widehat{X}_{\tau}$.
The proof is by induction on $\tau$. For $\tau = 1$, 
\[ 
\widehat{F}_{1} \equiv F_1 \quad\mbox{and}\quad \widehat{X}_{1} \equiv X_1
\]
Moreover, 
\[
\Prob{\neg \widehat{F}_{0} \wedge \widehat{X}_{0}} = 1
\]
Thus, from Lemma~\ref{lem::expansion} we have 
\[
\Prob{A_{1}} \geqslant 1 - \frac{1}{n^2}
\]
Now, assume that $\tau \geqslant 2$. It holds that
 \begin{eqnarray}\label{eq::atau}
\Prob{A_{\tau}} & = & \Prob{F_\tau \vee \widehat{F}_{\tau - 1} \vee (X_\tau \wedge \widehat{X}_{\tau})} \nonumber \\
& \geqslant & \Prob{(\widehat{F}_{\tau} \vee X_{\tau}) \wedge A_{\tau - 1}} \nonumber \\
& = & \Prob{\widehat{F}_{\tau} \vee X_{\tau} \;\left|\; A_{\tau - 1}}\Prob{A_{\tau - 1}\right.} \nonumber \\
& \geqslant & \left(1 - \frac{\tau - 1}{n^2}\right)\Prob{\widehat{F}_{\tau} \vee X_{\tau} \;\left|\;A_{\tau - 1}\right.} 
\end{eqnarray}

\noindent
where the last inequality is due to the induction hypothesis. We have that
\begin{eqnarray} \label{eq::f or r}
\Prob{\widehat{F}_{\tau} \vee X_{\tau} \;\left|\; A_{\tau - 1}\right.} & = &
\Prob{\widehat{F}_{\tau} \vee X_{\tau} \;\left|\; A_{\tau - 1}\wedge 
\widehat{F}_{\tau - 1}\right.}\Prob{\widehat{F}_{\tau - 1}\;\left|\; A_{\tau - 1}\right.} + \nonumber \\
& & \quad + \Prob{\widehat{F}_{\tau} \vee X_{\tau} 
\;\left|\; A_{\tau - 1}\wedge \neg\widehat{F}_{\tau - 1}\right.}\Prob{\neg\widehat{F}_{\tau - 1}\;\left|\; A_{\tau - 1}\right.} 
\end{eqnarray}
Since $A_{\tau - 1}\wedge \widehat{F}_{\tau - 1} \Rightarrow \widehat{F}_{\tau} \Rightarrow \widehat{F}_{\tau} \vee X_\tau$,
it holds that 
\begin{equation} \label{eq:: F or X 1}
\Prob{\widehat{F}_{\tau} \vee X_{\tau} \;\left|\; A_{\tau - 1}\wedge \widehat{F}_{\tau - 1}\right.} = 1
\end{equation}
Since $A_{\tau - 1}\wedge \neg\widehat{F}_{\tau - 1} \equiv \widehat{X}_{\tau - 1} \wedge \neg\widehat{F}_{\tau - 1}$ and
$(\widehat{F}_{\tau} \vee X_{\tau}) \wedge\widehat{X}_{\tau - 1} \wedge \neg\widehat{F}_{\tau - 1} \equiv (F_\tau \vee X_\tau) 
\wedge \widehat{X}_{\tau - 1} \wedge \neg\widehat{F}_{\tau - 1}$, it holds that
\begin{equation}\label{eq:: F or X 2}
\Prob{\widehat{F}_{\tau} \vee X_{\tau} \;\left|\; A_{\tau - 1}\wedge \neg\widehat{F}_{\tau - 1}\right.}
= \ \Prob{F_\tau \vee X_\tau \;\left|\; \neg\widehat{F}_{\tau - 1} \wedge \widehat{X}_{\tau - 1} \right.}
\geqslant  1 - \frac 1 {n^2} 
\end{equation}
where the last inequality follows from Lemma~\ref{lem::expansion}. By combining Ineq.s~\ref{eq::f or r}, \ref{eq:: F or X 1}, and \ref{eq:: F or X 2}, we obtain
\[
\Prob{\widehat{F}_{\tau} \vee X_{\tau} \;\left|\; A_{\tau - 1} \right.}
\geqslant \Prob{ \widehat{F}_{\tau -1 } \left|\; A_{\tau - 1} \right.   } + 
\left( 1- \frac 1 {n^2} \right)
\Prob{\neg  \widehat{F}_{\tau -1 } \left|\; A_{\tau - 1} \right.   }  \geqslant 1 -\frac 1 {n^2}  
\]
From Ineq. \ref{eq::atau}, we obtain
\[
\Prob{A_{\tau}} \geqslant  \left(1 - \frac{\tau - 1}{n^2}\right) \left( 1 -\frac 1 {n^2}   \right) \geqslant 1 - \frac{\tau}{n^2}
\]
\qed
 
\begin{lemma} \label{lm::IIphasetime}
Let $\mathcal{H}=\mathcal{H}(n,p,q,\alpha,\gamma)$ be a Home-MEG and assume that $\lceil \frac {5 \Lambda}{n} \rceil \leqslant  \min\{ 1/\alpha , 1/(4q) \}$. Then
\begin{equation*} 
\Prob{ \exists \ \tau  \leqslant      \frac{\log n}{\log K}  \ :  \   |C_\tau | \geqslant  \frac n{16}         } 
 \ \geqslant 1 -  \frac {\log n} {n^2}
\end{equation*}
\end{lemma}
\proof
We observe that for any  $\tau \geqslant \frac{\log (n/\log n)}{\log K} + 1$, event $\widehat{X}_{\tau }$ does not hold.
Then, for $\tau = \lfloor \frac{\log n}{\log k} \rfloor $, Lemma~\ref{lem::expansion2} implies
\[
\Prob{\widehat{F}_{\tau }}  =  \Prob{\widehat{F}_{\tau }  \vee \widehat{X}_{\tau }} \geqslant 
 1 - \frac{\tau}{n^2} \geqslant 1 - \frac{\log n}{ n^2}
\]
\qed

\noindent
Thanks to the above lemma, we can now bound the time of  Phase 2.

\begin{lemma} \label{lm::IIphasetimeBIS}
Let $\mathcal{H}=\mathcal{H}(n,p,q,\alpha,\gamma)$ be a Home-MEG and assume that $\lceil \frac {5 \Lambda}{n} \rceil \leqslant  \min\{ 1/\alpha , 1/(4q) \}$.
Then the time of Phase 2 is w.h.p. $\mathcal{O}\left( \frac{\log n}{\log K} \left\lceil \frac {5\Lambda }{n}\right\rceil\right)$.
\end{lemma}

\smallskip\noindent
\textbf{Phase 3.} During Phase 2, we have that, w.h.p.,  a time period $\hat \tau$ exists
such that 
\begin{equation} \label{eq::n.nodeIII}
|C_{\hat \tau}|  \ \geqslant \   n/16  
\end{equation}

\noindent
Now, let's look at the nodes in $C_{\hat \tau}$ and at the edges from such nodes to a node  $v \in V \setminus \cup_{0}^{\hat\tau -1}C_{i}$.
We observe such ``new'' edges for $\ell$ consecutive time steps after the end of time period $\hat \tau$.
Define binary r.v. $Y_v = 1 $ iff some of such edges is on in at least one of these $\ell$ time steps.
We make use of Lemma  \ref{lm::dtimeapx}    and  Ineq. \ref{eq::n.nodeIII} to derive  
\[
\Prob{Y_v =0} \ \leqslant \ \left( 1 - \frac{\ell}{\Lambda}\right)^{n/16}
\leqslant  \exp{ \left( - \frac{  \ell   n}{ 16 \Lambda }\right)}
\]

\noindent By choosing $\ell \geqslant    \frac{ 32 \Lambda  \log n}{  n}$, and applying the union bound
over all $Y_v$'s, we have that, w.h.p., all nodes will be connected to some node in $C_{\hat \tau}$ within the next  $\ell$ time steps.
We thus have proved that

\begin{lemma} \label{lm::IIIphasetime}
Let $\mathcal{H}=\mathcal{H}(n,p,q,\alpha,\gamma)$ be a Home-MEG and 
assume $\lceil    \frac {5 \Lambda }{   n  } \rceil \leqslant  \min\{ 1/\alpha , 1/(4q) \}$.
Then,  w.h.p. the time of Phase 3 is upper bounded by  
\begin{equation*}
  \mathcal{O}\left(  \frac {\Lambda \log n}{   n}   \right)
\end{equation*}
\end{lemma}

\noindent
Finally, Theorem~\ref{thm::flotime} follows from the time bounds of the three phases, i.e. Lemmas \ref{lm::boot}, \ref{lm::IIphasetimeBIS}, and \ref{lm::IIIphasetime}.

%% file: trunk/conclusions.tex
In this paper, we have introduced a simple opportunistic link model which is able to faithfully reproduce the power law+exponential tail dichotomy of inter-contact time typical of human mobility, and used this model to derive asymptotic bounds on flooding time in opportunistic networks. A major finding of our study is that flooding time in opportunistic networks might be much faster than what predicted by existing bounds based on simple  geometric mobility models which are known not to be able to reproduce the above described dichotomy. 

Our bounds on flooding time -- and, more specifically, the proof of Theorem \ref{thm::flotime}, which breaks down the flooding process in phases of different duration -- could  be very useful in estimating the propagation of proximity malware (i.e., malware propagated through Bluetooth/WiFi interfaces) in mobile networks \cite{Wang}. For instance, it has been recently observed in \cite{Zhu} that MMS malware (i.e., malware propagated through sending MMSs) in cellular networks displays a slow start followed by an exponential propagation. Our results can be used to investigate whether a similar propagation pattern is displayed also in proximity malware.

A major avenue for further improving our results is introducing dependency between the Markov chains modeling the possible links in the network, in order to account for physical mobility constraints and/or social relationships between nodes. Deriving bounds on flooding time with dependent pair-wise links is a challenging task, which we are currently undertaking.